\documentclass[12pt]{spieman}  
\usepackage{amsmath,amsfonts,amssymb}
\usepackage{graphicx}

\definecolor{updatecolor}{rgb}{0.5,0.0,0.8}
\newcommand{\updated}[1]{{#1}}

\title{STROBE-X Mission Overview}

\author[a,*]{Paul S. Ray}
\author[b]{Peter W. A. Roming}
\author[k]{Andrea Argan}
\author[c]{Zaven Arzoumanian}
\author[$\spadesuit$]{\updated{David R. Ballantyne}}
\author[d]{Slavko Bogdanov}
\author[e]{Valter Bonvicini}
\author[f]{Terri J. Brandt}
\author[g]{Michal Bursa}
\author[h]{Edward M. Cackett}
\author[i]{Deepto Chakrabarty}
\author[a]{Marc Christophersen}
\author[z]{Kathleen M. Coderre}
\author[j,\dag,\S]{Gianluigi De Geronimo}
\author[k]{Ettore Del Monte}
\author[k]{Alessandra DeRosa}
\author[z]{Harley R. Dietz}
\author[k]{Yuri Evangelista}
\author[k]{Marco Feroci}
\author[b]{Jeremy J. Ford}
\author[b]{Cynthia Froning}
\author[l]{Christopher L. Fryer}
\author[c]{Keith C. Gendreau}
\author[m]{Adam Goldstein}
\author[n]{Anthony H. Gonzalez}
\author[$\clubsuit$]{\updated{Dieter Hartmann}}
\author[o]{Margarita Hernanz}
\author[a]{Anthony Hutcheson}
\author[f]{Jean in ‘t Zand}
\author[p]{Peter Jenke}
\author[q]{Jamie Kennea}
\author[l]{Nicole M. Lloyd-Ronning}
\author[r]{Thomas J. Maccarone}
\author[$\ddag$]{Dominic Maes}
\author[c]{Craig B. Markwardt}
\author[s]{Malgorzata Michalska}
\author[c]{Takashi Okajima}
\author[o]{Alessandro Patruno}
\author[b]{Steven C. Persyn}
\author[b]{Mark L. Phillips}
\author[t]{Chanda Prescod-Weinstein}
\author[b]{Jillian A. Redfern}
\author[i]{Ronald A. Remillard}
\author[v]{Andrea Santangelo}
\author[b]{Carl L. Schwendeman}
\author[a]{Clio Sleator}
\author[u]{James Steiner}
\author[c]{Tod E. Strohmayer}
\author[g]{Jiri Svoboda}
\author[v]{Christoph Tenzer}
\author[b]{Steven P. Thompson}
\author[z]{Richard W. Warwick}
\author[w]{Anna L.~Watts}
\author[x]{Colleen A.~Wilson-Hodge}
\author[y]{Xin Wu}
\author[a]{Eric A. Wulf}
\author[e]{Gianluigi Zampa}

\affil[a]{Space Science Division, U.S. Naval Research Laboratory, Washington, DC 20375 USA}
\affil[b]{Southwest Research Institute, San Antonio, TX 78238 USA}
\affil[c]{NASA's Goddard Space Flight Center, Greenbelt, MD 20771 USA}
\affil[d]{Columbia Astrophysics Laboratory, Columbia University, New York, NY 10027 USA}
\affil[e]{Istituto Nazionale di Fisica Nucleare (INFN), Trieste TS, Italy}
\affil[f]{SRON Netherlands Institute for Space Research, Niels Bohrweg 4, 2333 CA Leiden, The Netherlands}
\affil[g]{Astronomical Institute of the Czech Academy of Sciences, Bo\v{c}n\'{i} II 1401, 14100 Prague, Czech Republic}
\affil[h]{Department of Physics \& Astronomy, Wayne State University, Detroit, MI 48201, USA}
\affil[i]{Kavli Institute for Astrophysics and Space Research, Massachusetts Institute of Technology, Cambridge, MA 02139 USA}
\affil[j]{Department of Nuclear Engineering and Radiological Sciences, Univeristy of Michigan, Ann Arbor, MI, USA}
\affil[k]{INAF - Istituto di Astrofisica e Planetologia Spaziali, 00133, Roma, ITALY\\}
\affil[l]{Los Alamos National Laboratory, Los Alamos, NM, United States, 87545 USA}
\affil[m]{Universities Space Research Association, Huntsville, AL 35805, USA}
\affil[n]{Department of Astronomy, University of Florida, Gainesville, FL 32611, USA}
\affil[o]{Institute of Space Sciences (ICE-CSIC) and  IEEC, Campus UAB, Carrer de Can Magrans s/n, 08193 Bellaterra (Barcelona), Spain}
\affil[p]{University of Alabama in Huntsville, Huntsville, AL 35805, USA}
\affil[q]{Department of Astronomy and Astrophysics, The Pennsylvania State University, University Park, PA 16802, USA}
\affil[r]{Department of Physics \& Astronomy, Texas Tech University, Lubbock, TX, 79409, USA}
\affil[s]{Space Research Center, Polish Academy of Sciences, Warszawa, Poland}
\affil[t]{Department of Physics and Astronomy, University of New Hampshire, Durham, NH 03824 USA}
\affil[u]{Harvard-Smithsonian Center for Astrophysics, Cambridge, MA 02138 USA}
\affil[v]{Eberhard Karls Universit\"at, T\"ubingen, GERMANY}
\affil[w]{Anton Pannekoek Institute for Astronomy, University of Amsterdam, 1090GE Amsterdam, The Netherlands}
\affil[x]{Astrophysics Branch, NASA's Marshall Space Flight Center, Huntsville, AL 35812, USA}
\affil[y]{Department of Nuclear and Particle Physics, University of Geneva, Geneva, CH, Switzerland}
\affil[z]{Lockheed Martin Co.}
\affil[$\ddag$]{BAE Systems Space \& Mission Systems Inc., Boulder, CO 80301, USA}
\affil[$\dag$]{Stony Brook University, Electrical and Computer Engineering, Stony Brook, NY 11794, USA}
\affil[$\S$]{DG Circuits, dgcircuits.com, Syosset, NY 11973, USA}
\affil[$\clubsuit$]{Department of Physics and Astronomy, Clemson University, Clemson, SC 29634, USA}
\affil[$\spadesuit$]{Georgia Institute of Technology, Atlanta, GA 30332, USA}
\begin{document} 
\maketitle

\begin{abstract}
We give an overview of the science objectives and mission design of the \textit{Spectroscopic Time-Resolving Observatory for Broadband Energy X-rays (STROBE-X)} observatory, which has been proposed as a NASA probe-class ($\sim$\$1.5B) mission in response to the Astro2020 recommendation for an X-ray probe. 
\end{abstract}


{\noindent \footnotesize\textbf{*}\linkable{paul.s.ray3.civ@us.navy.mil} }


\section{Introduction}
\label{sect:intro}  

We present an overview of the \textit{Spectroscopic Time-Resolving Observatory for Broadband Energy X-rays (STROBE-X)} mission, which has been proposed in response to NASA's 2023 call for probe-class X-ray missions. The Astrophysics Probe (APPROBE) call was motivated by the Decadal Survey on Astronomy and Astrophysics 2020 (Astro2020)\cite{Astro2020}, which made a competed line of probe-class (cost cap $\sim$ \$1.5B) space missions one of its top priorities. The first Probe will be either an X-ray or Far-IR mission, with the Probe line continuing at a cadence of about one mission per decade. In addition, Astro2020's highest priority recommendation for space was a Time Domain Astrophysics Program that would ``expand space-based time-domain and [multi-messenger] follow-up facilities in space''. With its focus on studying the time-variable X-ray sky with unprecedented capabilities, STROBE-X addresses both of these recommendations and will serve the astrophysics community as the X-ray component of the multi-wavelength and multi-messenger facilities in the 2030s.

The mission draws together well-developed technologies and instrument concepts. The primary pointed instrument (LEMA, see \S\ref{sec:lema}) is a scaled-up version of the NICER X-ray Timing Instrument \updated{(XTI)} currently operating on the International Space Station\cite{2016SPIE.9905E..1HG}. The higher-energy pointed instrument (HEMA, see \S\ref{sec:hema}) and the wide-field monitor (WFM, see \S\ref{sec:wfm}) have been developed for over a decade for the  LOFT and eXTP concepts \cite{2014SPIE.9144E..2WZ,2014SPIE.9144E..2VB,2022SPIE12181E..1XF,2018SPIE10699E..48H}. The merged STROBE-X mission concept was proposed in 2016 and then selected for study by NASA in 2017, as part of the preparation for the Astro2020 survey.  The study report\cite{2019arXiv190303035R}, along with the other candidate Probe missions, was provided as input to the Astro2020 committee.

\section{Science Overview}

\begin{figure}[htb]
    \begin{center}
        \includegraphics[width=4.5in]{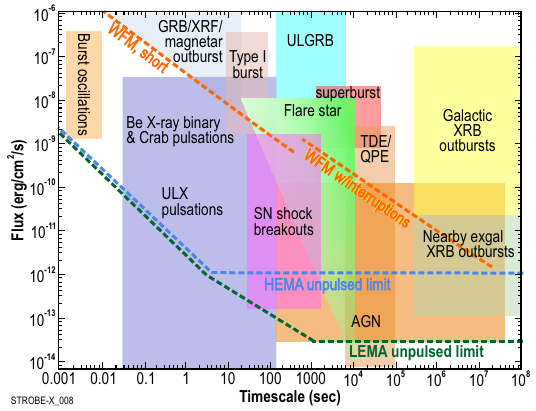}
    \end{center}
    \caption{STROBE-X provides the X-ray eyes on the dynamic universe, observing transient sources and phenomena spanning 8 orders of magnitude in flux and 12 orders of magnitude in timescale. \updated{The colored regions show the areas of Flux-Timescale phase space occupied by various classes of transients. The dashed lines show the sensitivity limits for the various instruments. The HEMA and LEMA lines become horizontal at the background systematics limit for unpulsed sources. For coherent pulsations, their sensitivity continues to improve with time indefinitely.}}
    \label{fig:dynamicrange}
\end{figure}

To advance the broad science impacted by astrophysical transient and variable sources, we must understand their underlying engines, and the key measurements lie in the X-ray band. X-ray data on transients have grown immensely, with a \updated{extensive archives} from multiple missions including Swift, Fermi, RXTE, and NICER. However, each of these missions lacked one or more critical capabilities: a large field-of-view (FoV), large effective area, broadband spectral coverage, and/or high time resolution. Because transients occur randomly across the sky, a large FoV is required to rapidly identify and localize them for detailed follow-up. Transients also span a large dynamic range in flux, timescale, and X-ray bandpass, requiring instrumentation that accommodates wide ranges in all three with sufficient sensitivity on all variability timescales (Fig. \ref{fig:dynamicrange}).  

STROBE-X is designed to obtain exactly these observations. STROBE-X is a pointed observatory that conducts \updated{broad energy band spectroscopy of time-variable targets using narrow FoV instruments, while simultaneously monitoring the full sky (with a wide FoV instrument) for transient sources}. It continuously monitors the dynamic X-ray sky to detect and localize new transients with the very large (4 sr) FoV Wide-Field Monitor (WFM; 2--50 keV). Immediately following localization, the rapidly slewing spacecraft (S/C) autonomously points to the transient within 7--24 min, enabling the pointed instruments---the 0.2-12 keV Low Energy Modular Array (LEMA; 8.5$\times$ more effective area than NICER) and the 2--30 keV High Energy Modular Array (HEMA; 5.5$\times$ more effective area and $\sim 3\times$ better energy resolution than RXTE/PCA)---to obtain high count rate, time-resolved, broadband X-ray spectroscopy. In between transient outbursts, WFM continuously measures the flux and spectral activity of thousands of X-ray sources across the sky, while LEMA and HEMA make pointed observations of scheduled sources. In addition to on-board transient detection, STROBE-X can respond rapidly to externally triggered targets of opportunity (ToO), starting autonomous repointing within 2 minutes of automated ground alerts and executing human-triggered ToOs within $<24$ hours.

STROBE-X can probe an enormous dynamic range in both timescales (from $\mu$s to years, with dense sampling, not just occasional visits) and fluxes (from faint pulsars to the brightest sources in the X-ray sky), making it extremely versatile and capable (see Fig. \ref{fig:dynamicrange}). With its huge collecting area, fast timing, good energy resolution, and broadband coverage, STROBE-X changes the paradigm of X-ray spectroscopy. With STROBE-X, we move from discrete time-averaged snapshots that smear out any time-evolution to dynamic spectroscopy: broadband spectral ``movies'' that reveal the real-time evolution of astrophysical phenomena, elucidating their underlying physics by resolving the relevant physical timescale (jet launching, recombination, state change, spin period, precession period, etc.). Instantaneous wide-field sensitivity coupled with high-sensitivity pointed instruments make STROBE-X the essential X-ray partner for high-cadence synoptic Time Domain and Multi-Messenger (TDAMM) investigations in other electromagnetic bands (radio, millimeter, infrared, optical, ground and space-based gamma-ray) as well in \updated{gravitational waves (GWs)} and high-energy neutrinos.  

\updated{The primary pointed instruments LEMA and HEMA are non-imaging --- a design choice that enables almost an order of magnitude increase in collecting area in a Probe class mission. The focus on transient and variable sources means that most targets will effectively be point sources and imaging of the narrow fields is not essential. Of course, source localization is critical and the all-sky imaging provided by the WFM gives source positions accurate to an arcminute, which is sufficient both to direct STROBE-X pointed observations and to allow single field observations with ground-based telescopes.}

The STROBE-X PI-led investigation will apply powerful capabilities to address four science goals:  
\begin{itemize}
\item{Multi-messenger: Detect and localize the X-ray counterparts of GW and neutrino sources to understand their progenitor systems, central engines, jets, compact remnants, and ejecta. }
\item{Time Domain Astronomy of Explosive Transients: Discover, follow up, and provide the X-ray characterization of new events to determine what powers a wide variety of explosive phenomena throughout our universe.  }
\item{Gravity \& Extreme Physics: Probe the extremes of relativistic and particle astrophysics by observations of black hole (BH) spins across the mass spectrum and \updated{measurements of} neutron star (NS) equation of state, surface configuration and magnetic field geometries. }
\item{The Dynamic Universe: Provide the X-ray view of the variable sky with wide-field monitoring and broadband spectroscopy. }
\end{itemize}

The STROBE-X mission also includes a robust general observer (GO) program that will encompass at least 70\% of the observing time, \updated{beginning after the commissioning phase, and interleaved in the schedule equally with the PI-led investigation targets}.  As a highly responsive and flexible X-ray observatory with transformational capabilities and a substantial GO time allocation, STROBE-X will serve a very large community with a broad portfolio of astrophysical applications, including: supernovae (SNe), gamma-ray bursts (GRBs), active galactic nuclei (AGN), tidal disruption events (TDEs), X-ray binary accretion physics, magnetized accretors, ultra-luminous X-ray sources (ULXs), stellar coronae, supernova remnants (SNRs), binary stellar evolution, isolated NS, clusters and groups of galaxies, and comets. Not included in this list are the new phenomena inevitably revealed when substantially new observational capabilities are enabled. An in-depth description of many of the specific science investigations that STROBE-X will undertake is presented in the study report\cite{2019arXiv190303035R}.

\section{Instrument Overview}

\begin{figure}[htbp]
    \begin{center}
        \includegraphics{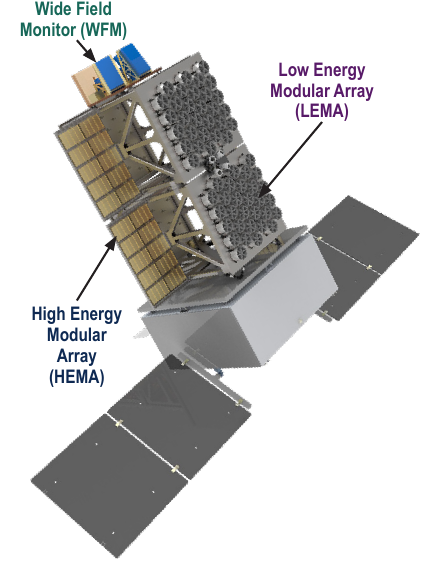}
    \end{center}
    \caption{Rendering of the STROBE-X observatory, showing the 3 instruments. \updated{Note that the pointing of LEMA and HEMA are co-aligned. The HEMA FoV is 1$^\circ$ so it is not obscured, even from its set-back position on the optical bench. This figure shows the location of the WFM on the spacecraft, while the detailed placement of the four camera pairs can be found in the accompanying paper on the WFM (Ref. \citenum{JATISWFM}, Figure 3 and \S5.1).}}
    \label{fig:overview}
\end{figure}

STROBE-X carries a suite of 3 instruments (see Fig. \ref{fig:overview}) that work together to provide situational awareness of the X-ray sky and characterize the broadband spectral and timing behavior of sources from 0.2--50 keV.

\subsection{Low-Energy Modular Array (LEMA)}
\label{sec:lema}
LEMA is the central instrument on STROBE-X. It is an array of 60 sets of concentrating optics paired with small commercial silicon drift detectors (SDDs) that efficiently collect X-rays in the 0.2--12 keV band. The optics and detectors are based on those currently operating on the NICER instrument, with the optics scaled up to a larger area and longer (3 m) focal length. The FoV is 4 arcmin (FWHM) and each photon is time-tagged to an accuracy of better than 200 ns with an energy resolution of 85 eV (at  1keV). Further instrument details are provided elsewhere in this volume\cite{JATISLEMA}.

An essential feature of LEMA is that it provides the optical bench that supports the entire instrument suite, with the HEMA panels attached to the sides and the WFM mounted on top \updated{(as shown in Figure \ref{fig:overview})}. The structure is constructed in two halves to increase redundancy, reduce the size of the assembly and test fixtures, and allow parallelization in assembly, integration, and test flow. Throughout, STROBE-X is designed with efficient integration flow and schedule flexibility.

\subsection{High-Energy Modular Array (HEMA)}
\label{sec:hema}
\begin{figure}[htbp]
\begin{center}
    \includegraphics[width=4.0in]{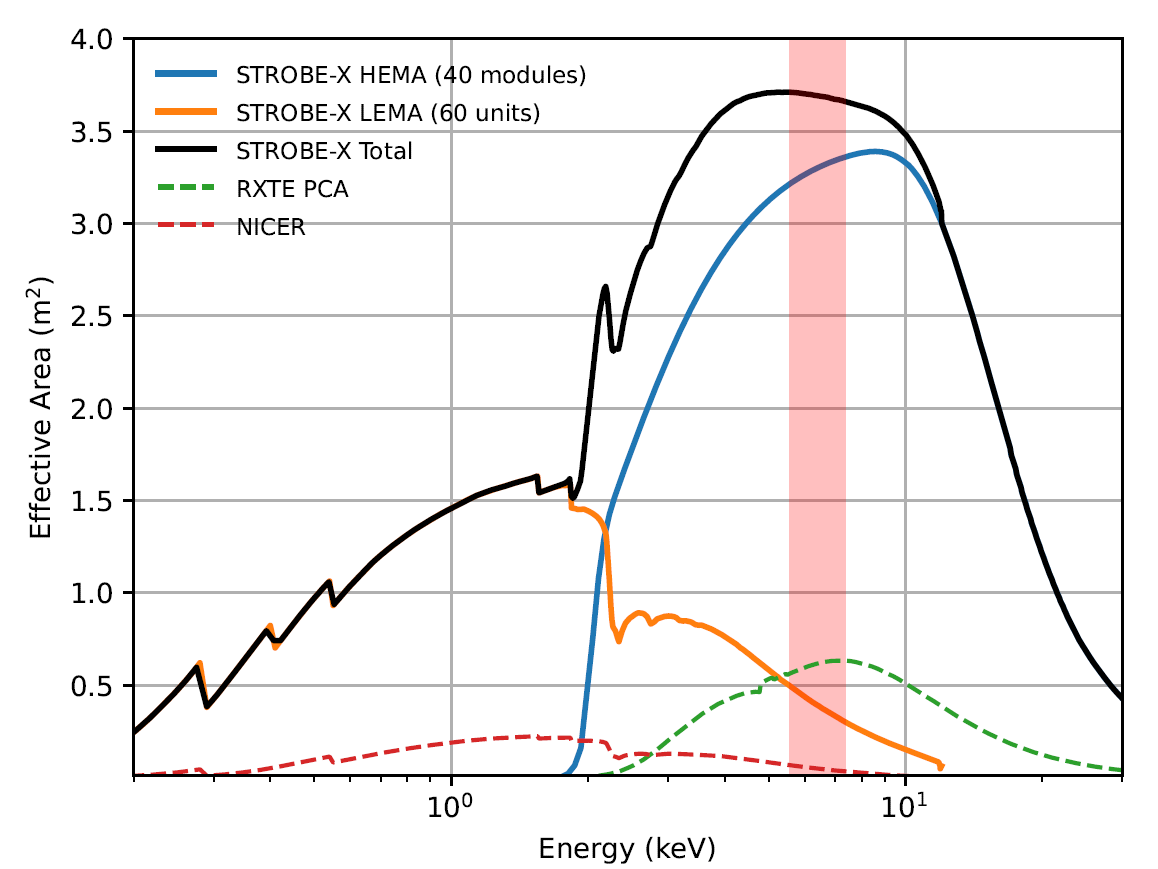}
  \end{center}
    \caption{STROBE-X effective area, showing LEMA, HEMA individually and the total of the pointed instruments. For comparison, NICER and RXTE -- the previous high-throughput missions in those energy bands -- are shown on the same scale, illustrating the transformational increase in collecting area achieved. The pink shaded bar highlights the critical Fe-K line region.}
    \label{fig:effarea}
\end{figure}

HEMA is a highly modular array of large-area SDDs that employ thin glass microcapillary plates to collimate 2--30 keV X-rays to a 1$^\circ$ FoV, co-aligned with the LEMA FoV. Complete details on HEMA are provided elsewhere in this volume\cite{JATISHEMA}. 
HEMA complements LEMA by providing a massive collecting area through the Fe-K region around 6--7 keV and above \updated{(see Figure \ref{fig:effarea})}, to characterize the non-thermal emission, reflection, reverberation, and its relationship to the lower energy emission observed with LEMA.

\subsection{Wide Field Monitor (WFM)}
\label{sec:wfm}
The WFM serves as the X-ray sentinel for STROBE-X. It comprises 4 coded-aperture mask camera pairs that employ the same SDD technology as HEMA, but with finer anode pitch to provide imaging. The nominal energy band is 2--50 keV, with an energy resolution of 300 eV. Each camera pair makes 2-D images with a half-coded FoV of $65^\circ \times 65^\circ$ combining to give an instantaneous FoV of 4 sr (32\% of the sky, \updated{which is almost 3$\times$ that of Swift/BAT}). The source localization accuracy is 1 arcmin.  A distinguishing feature is that all events are telemetered to the ground, giving maximum flexibility for analyses with high spectral and timing resolution.
Source detection and transient triggering are performed onboard, allowing the WFM to issue prompt burst alerts to the community and trigger autonomous repoint requests for the pointed instruments to allow them to get on the source within minutes of a burst (actual time depends on the length of slew). The technical details of the WFM are provided in a separate paper\cite{JATISWFM}.

\subsection{Instrument Summary Table}

\begin{table}[h]
    \caption{Instrument characteristics}
    \label{tab:inst}    
    \centering
    {\small 
    \begin{tabular}{l|ccc}
    \hline
    \textbf{Quantity} & \textbf{LEMA} & \textbf{HEMA} & \textbf{WFM} \\
    \hline
    Bandpass  & 0.2--12 keV & 2--30 keV & 2--50 keV \\
    Effective Area & 16,000 cm$^2$ &  34,000 cm$^2$ & 364 cm$^2$ \\
    Energy Resolution & 85 eV (@ 1keV) & $<$300--500 eV (@ 6.4 keV) & 300 eV (@6.4 keV) \\
    Number of Units & 60 XRCs & 40 modules & 4 camera pairs \\
    FoV & 4$^\prime$ (FWHM) & 1$^\circ$ (FWHM) & 32\% of sky (half coded) \\
    Absolute Time Accuracy & 200 ns & 7 $\mu$s & 24 $\mu$s \\
    Count Rate on 1 Crab & 111,000 c/s & 104,000 c/s & 380 c/s \\
    \hline
    \end{tabular}
}
\end{table}

\section{Mission Design and Spacecraft}

The STROBE-X instrument suite is hosted on a proven, agile spacecraft platform built by Lockheed Martin. \updated{The bus is based on over four decades of LM spacecraft heritage, using proven subsystems and architecture from most recent missions as OSIRIS-REx, MAVEN and MRO as well as leveraging in-development designs from DAVINCI and VERITAS.}

The observatory operates in a circular low-inclination orbit ($<15^\circ$ inclination) and uses a propulsion system to keep the orbit altitude between 550 km and 575 km, \updated{as well as to perform any needed collision avoidance maneuvers}. Attitude control is provided by high-performance reaction wheels and thrusters, based on sensor inputs from star trackers, an inertial reference unit, and a sun sensor. Attitude knowledge is better than 20 arcsec with control better than 60 arcsec.  The attitude control systems allow the spacecraft to slew 45$^\circ$ in 7.6 min and 90$^\circ$ in 11.5 min. This allows the pointed instruments to observe at least 3 targets per orbit and avoid observing inefficiencies caused by Earth occultations. \updated{The overall observing efficiency is about 65\% with the nominal slewing performance, based on a detail week-in-the-life simulation (Figure \ref{fig:dayinthelife} shows a single day).}
 A GPS receiver provides position, velocity and time knowledge to better than 100 ns.
The spacecraft can point the  HEMA and LEMA instruments anywhere in the sky outside of a 45$^\circ$ exclusion zone around the Sun.

\begin{table}[h]
    \caption{Observatory and Mission characteristics}
    \label{tab:mission}    
    \centering
    \begin{tabular}{lc}
    \hline
    \textbf{Quantity} & \textbf{Value} \\
    \hline
    Dry Mass & 3519 kg (MEV) \\
    Power (Science Mode) & 2170 W (MEV) \\
    Orbit & LEO, 550--575 km, $<15^\circ$ inclination \\
    Pointing Control (3-axis, 3$\sigma$) & 39$^{\prime\prime}$ (14$^{\prime\prime}$ knowledge) \\
    Sun Avoidance Angle & 45$^\circ$ \\
    Prime Mission Duration & 5 years \\
    Deployments & Solar array deploy, HGA boom deploy \\
    Science Data Volume & 348 Gb/day \\
    \end{tabular}

\end{table}

\subsection{Communications}

The primary telecommunications are done using Atlas Space commercial antenna networks, which uplink commands and downlink telemetry using S-band low-gain antennas (LGA). The primary high-rate data downlink makes use of a Ka-band link with a gimballed high-gain antenna (HGA). This system can achieve 150 Mbps downlink speeds during contacts, sufficient to get the event data down promptly. For the nominal daily data volume, approximately 7 contacts per day are required. An onboard solid state recorder (with a capacity of about a week of science data) stores the data until it can be downlinked. 

A critical feature of a time-domain mission like STROBE-X is the ability to broadcast alerts to the community and receive target of opportunity commands from the ground with minimal latency. With TDRSS being decommissioned, missions can no longer rely on their demand access feature to provide low-latency commanding and telemetry. A replacement has yet to be defined by NASA, so the baseline plan for STROBE-X is to employ a commercial solution of L-band space-to-space links from Inmarsat. This allows onboard burst alerts to be relayed to the ground for distribution to the community in $<5$ minutes. In addition, ground alerts from other facilities or multi-messenger triggers can be sent up to the spacecraft.

\subsection{Mission Operations}

\begin{figure}[htbp]
    \includegraphics[width=6.0in]{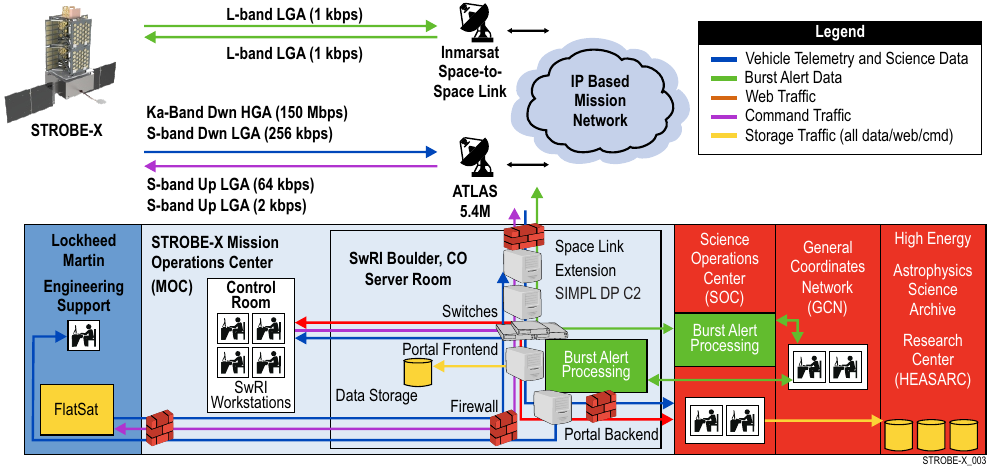}
    \caption{Mission Operations Architecture}
    \label{fig:MOC}
\end{figure}

The STROBE-X Mission Operations Center (MOC) is hosted by SwRI in Boulder, Colorado. The mission operations architecture connects the MOC with the ground network, space-to-space network,
Science Operations Center (SOC), General Coordinates Network (GCN), and spacecraft factory (see Fig. \ref{fig:MOC}).

The MOC is responsible for operating the spacecraft and science instruments. SwRI’s MOC is designed as a multi-mission,
multi-S/C facility and meets all requirements
necessary to implement the STROBE-X concept
of operations. The MOC has built-in redundancy
and \updated{is} certified to meet NASA IT security
standards. The MOC is staffed 8 hours a day,
5 days a week for baseline science operations
but has necessary facilities to support 24/7
operations during STROBE-X commissioning,
critical, and anomaly operations. The MOC is
designed to operate autonomously for up to 72
hours. Any problem requiring Flight Control
Team (FCT) attention triggers an alert to on-call
personnel who can review system data remotely
to determine if further anomaly resolution at the
MOC is necessary. The MOC also has Lockheed Martin factory
reach-back for S/C bus telemetry trending and
performance monitoring (on a monthly basis)
and anomaly resolution, as needed.

\subsection{Science Operations}

STROBE-X Science Operations will work on a model that was pioneered by the Neil Gehrels Swift Observatory, with improvements to make STROBE-X more responsive to alerts from other missions. Science Operations will need to respond to both follow-up of WFM triggers and Target of Opportunity requests both from members of the community and automated requests generated from external triggering instruments such as other GRB missions, Neutrino and GW observatories.

The Science Operations Center will operate during business hours\updated{, with a single Observatory Duty Scientist (ODS) on call to handle out of hours issues. Observation plans for STROBE-X are created from an initial long-term plan of observations, with updates from recently received TOO requests and WFM triggers. STROBE-X Observing plans are discussed, created and delivered daily within a 24 hour period on weekdays, with longer plans created to cover the weekend. This rapid planning process ensures that STROBE-X can respond to transients quickly.}

Faster response is handled by a combination of automation and \updated{ODS response. ODS will manually handle any TOO requests for any non-immediate TOOs. Immediate TOO requests will be handled by an automated system that with low-latency assesses feasibility and safety of a TOO request, and creates a command upload. This immediate response will allow STROBE-X to respond to external triggers with almost as low latency as onboard WFM triggers. The ODS is paged 24/7 for WFM triggers, immediate TOOs, and regular TOO requests, ensuring that at least one SOC member is aware of what STROBE-X's status at all times.}

\begin{figure}[htbp]
    \includegraphics[width=6.0in]{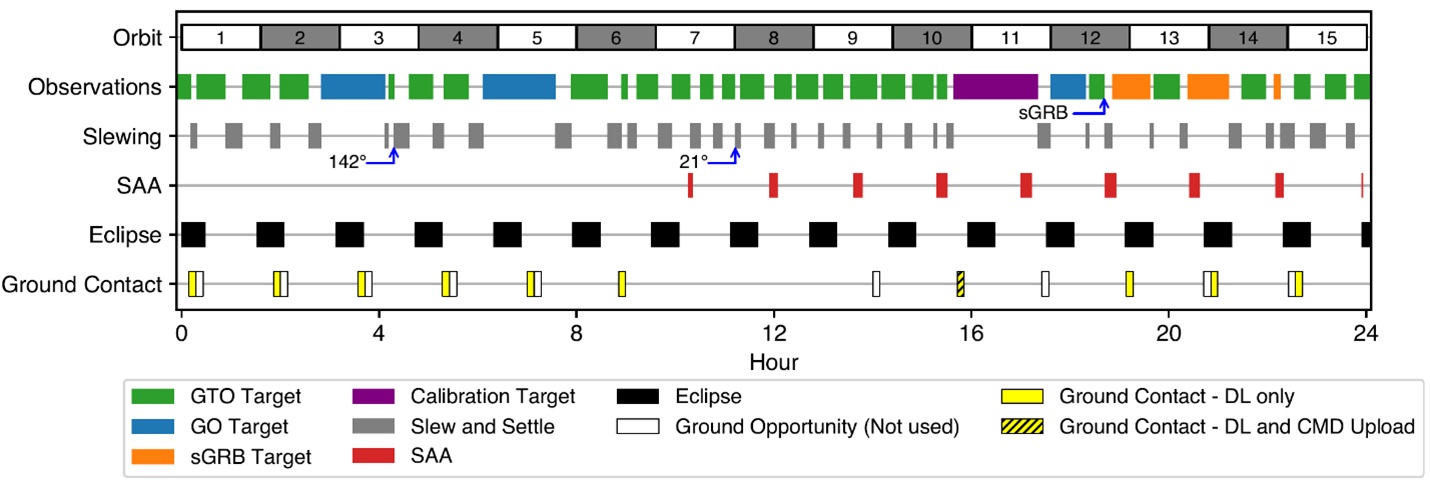}
   \caption{STROBE-X's typical day-in-the-life timeline. Observations show time spent pointing at \updated{Guaranteed Time Observer (GTO)} targets (green), GO targets (blue), calibration targets (purple), and GRB observations as a result of an Autonomous Repoint Request (orange) from a WFM GRB trigger (arrow). Time spent slewing in SAA and in eclipse and ground contacts for stations in Ghana and Rwanda (with their utilization) are also shown.}
    \label{fig:dayinthelife}
\end{figure}

It is noted that Swift has already demonstrated this capability for a Fast Radio Burst follow-up program using a continuous TDRSS Forward link. The improvements in communications for STROBE-X will carry this heritage from Swift, and allow STROBE-X to be a truly next-generation responsive TDAMM mission.  \updated{A sample day-in-the-life timeline is shown in Figure \ref{fig:dayinthelife}.}

\section{Summary}
\updated{STROBE-X will be a TDAMM observatory and an unparalleled laboratory for probing strong gravity and the behavior of matter in extreme conditions. The mission monitors the sky with a wide field monitor of unprecedented capability, providing situational awareness and alerts both for the community and for targeting the pointed instruments. The agile observatory can get the pointed instruments rapidly on source anywhere in the sky outside the 45$^\circ$ solar exclusion zone, where they will make spectral-timing measurements with transformational sensitivity. STROBE-X will measure the fundamental physical parameters of compact objects across the mass spectrum, revealing how they form, grow, and die, and will be the critical X-ray partner in the new era of time-domain surveys. }

\updated{This paper presented the overall mission architecture, while companion papers present the technical details of the instrument designs of LEMA \cite{JATISLEMA}, HEMA \cite{JATISHEMA}, WFM \cite{JATISWFM}, and the ASIC readout electronics used for both HEMA and WFM \cite{JATISASIC}.
}

\appendix    

\subsection*{Disclosures}
The authors have no potential conflicts of interest to disclose.

\subsection* {Code, Data, and Materials Availability} 
Data sharing is not applicable to this article, as no new data were created or analyzed.

\subsection*{Acknowledgments}

Portions of this work performed at NRL were supported by NASA and ONR 6.1 basic research funding.
Italian authors acknowledge support by the Italian Space Agency through grant 2020-3-HH.1. J.S. acknowledges PRODEX PEA 4000144379, GACR project 21-06825X, and institutional support from RVO:67985815. A.L.W. acknowledges support from ERC Consolidator Grant No.~865768 AEONS.


\bibliography{report}   
\bibliographystyle{spiejour}   


\vspace{2ex}\noindent\textbf{Paul Ray} is an astrophysicist at the U.S. Naval Research Laboratory. He received his A.B. in physics from the University of California, Berkeley in 1989 and his Ph.D. in physics from Caltech in 1995. His current research interests include multiwavelength studies of pulsars and other neutron star systems and X-ray instrument and mission development. He is a member of AAS, IAU, and Sigma Xi.

\vspace{2ex}\noindent\textbf{Pete Roming} is Director of Space Engineering at Southwest Research Institute, and PI of the STROBE-X mission. He received his Ph.D. in physics and astronomy from Brigham Young University in 1998.  He is a certified Project Managment Professional and is PI of the UVOT instrument on Swift.

\vspace{1ex}
\noindent Biographies and photographs of the other authors are not available.


\clearpage

\end{document}